\documentclass{bmvc2k}

\usepackage{graphicx}
\usepackage{booktabs}
\usepackage{multirow}
\usepackage{makecell}
\usepackage{xcolor}
\usepackage{amsmath,amssymb}

\title{ML-CLIPSim: Multi-Layer CLIP Similarity for Machine-Oriented Image Quality}
\addauthor{Feng Ding}{feng_ding@sfu.ca}{1}
\addauthor{Haisheng Fu}{Haisheng.fu@ubc.ca}{2}
\addauthor{Jie Liang$^*$}{jliang@eitech.edu.cn}{3}
\addauthor{Qihan Xu}{qxa18@sfu.ca}{1}
\addauthor{Siyu Zhu}{zhusiyu@stu.xjtu.edu.cn}{4}
\addauthor{Jingning Han}{jingning@google.com}{5}

\addinstitution{
 Simon Fraser University\\
 Burnaby, BC, Canada
}
\addinstitution{
 University of British Columbia\\
 Vancouver, BC, Canada
}
\addinstitution{
 Eastern Institute of Technology\\
 Ningbo, Zhejiang, China
}
\addinstitution{
 Xi’an Jiaotong University\\
 Xi’an, Shaanxi, China
}
\addinstitution{
 Google Inc\\
 Mountain View, CA, USA
}
\runninghead{Student, Prof, Collaborator}{BMVC Author Guidelines}

\begin{document}

\maketitle

\noindent\textsuperscript{*}Corresponding author.
%-------------------------------------------------------------------------
\begin{abstract}
Image quality assessment has traditionally been defined in terms of signal fidelity or agreement with human perception. However, in many contemporary vision pipelines, images are ultimately consumed by downstream machine models, including classifiers, detectors, segmenters, and vision-language models. In such settings, a desirable quality measure should reflect not only visual appearance, but also the preservation of information that supports stable machine inference. This paper studies full-reference image quality from this machine-centric perspective. We formulate machine-oriented image quality as a latent machine utility, which measures the extent to which a distorted image preserves prediction-relevant information across a population of downstream models. Since this utility cannot be directly observed, we approximate it through pairwise predictive-consistency comparisons. Specifically, we construct the Predictive Consistency Dataset for Machine Perception (PCMP), where PSNR-matched distortion pairs are assigned soft preference labels by aggregating consistency votes from multiple pretrained models. Based on this supervision, we propose Multi-layer CLIP Similarity (ML-CLIPSim), a differentiable full-reference quality metric built upon a frozen CLIP visual encoder. ML-CLIPSim learns lightweight aggregation over intermediate patch-token representations and global image embeddings, allowing it to capture both localized evidence degradation and high-level semantic consistency. Experiments on the proposed PCMP dataset, external machine-preference benchmarks, human-IQA datasets, and learned image compression tasks show that ML-CLIPSim better aligns with machine-oriented preferences than conventional fidelity and perceptual metrics, while maintaining competitive correlation with human quality judgments. When used as a distortion term for learned image compression, ML-CLIPSim further improves rate--task trade-offs across multiple downstream tasks and codec architectures.
\end{abstract}

\section{Introduction}
\begin{figure}[t]
\centering
\includegraphics[width=\linewidth, trim=70 160 50 140, clip]{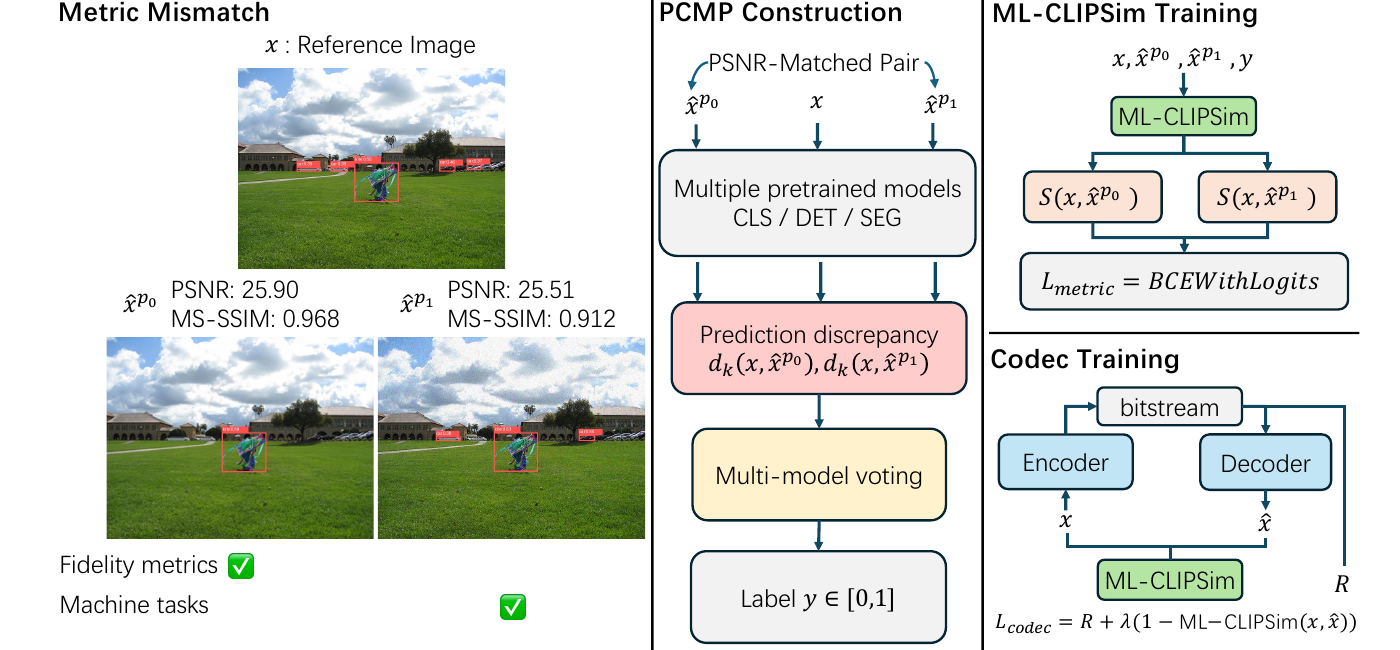}
\caption{
Overview of the proposed framework.
Left: fidelity metrics do not always reflect downstream-model behavior.
Middle: PCMP constructs PSNR-matched distortion pairs with soft labels from multi-model predictive consistency.
Right: ML-CLIPSim is trained from these pairwise preferences and used for learned image compression.
}
\label{fig:framework}
\end{figure}

Image quality assessment (IQA) has long been studied as a problem of measuring signal fidelity or perceptual agreement with human visual judgments. Classical full-reference metrics, such as PSNR, SSIM, and MS-SSIM~\cite{wang2004image,wang2003multiscale}, quantify pixel-level or structural distortions, while more recent learned metrics, such as LPIPS and DISTS~\cite{zhang2018unreasonable,ding2020image}, improve the correlation with human perception by comparing deep feature representations. These metrics have been highly successful when the primary consumer of an image is a human observer. However, this assumption is increasingly mismatched with modern visual communication and analysis pipelines, where images are often transmitted, stored, compressed, and ultimately processed by downstream machine-vision models, including classifiers, detectors, segmenters, and vision-language models.

In machine-oriented scenarios, distortions with similar fidelity scores may lead to substantially different downstream predictions, while visually noticeable distortions may have little effect on machine inference if task-relevant evidence is preserved. 
This mismatch motivates machine-oriented image quality assessment: machine-oriented quality measure should not merely ask whether a distorted image looks similar to its reference image, but whether it preserves the visual evidence that supports stable and reliable machine predictions. 
Fig.~\ref{fig:framework} illustrates this observation, where distortions with similar fidelity scores can lead to different downstream model behaviors, motivating the need for a machine-oriented quality measure.

A straightforward way to address this issue is to optimize images or codecs directly with downstream task losses, as in task-driven image coding and coding-for-machines frameworks~\cite{duan2020video,le2021image}. While effective for a specified model and task, such an approach has several practical limitations. It usually requires task annotations or pseudo-labels, depends on the architecture and loss function of a particular downstream model, and may not generalize well to unseen tasks, datasets, or model families. Moreover, directly incorporating multiple downstream models into image quality learning or compression optimization can be computationally expensive and difficult to scale. These limitations motivate the need for a task-agnostic yet machine-aware quality surrogate: a full-reference metric that can be trained from machine behavior, remains differentiable, and can be reused across downstream tasks without requiring task-specific supervision at deployment time.

To this end, we revisit image quality from a machine-centric perspective and formulate machine-oriented image quality as a \emph{latent machine utility}. This utility reflects how well a distorted image preserves prediction-relevant information across downstream models. Unlike human opinion scores or pixel-wise distortion values, this utility is not directly observable. It depends on heterogeneous models, tasks, and output spaces, whose prediction changes are not naturally calibrated on a common absolute scale. We therefore approximate this latent utility through pairwise predictive-consistency comparisons. Given two distorted versions of the same reference image, the one that induces smaller changes in downstream model predictions is considered to better preserve machine-relevant information.

Based on this formulation, we construct the \emph{Predictive Consistency Dataset for Machine Perception} (PCMP). A key challenge in constructing such supervision is to avoid learning a trivial preference for lower distortion magnitude. To reduce this confounding factor, PCMP compares only \emph{PSNR-matched} distortion pairs, so that the two candidates have similar low-level fidelity with respect to the reference image. Preference labels are then obtained by aggregating predictive-consistency votes from multiple pretrained models spanning classification, detection, and segmentation. The resulting soft labels reflect the degree of agreement among downstream models and provide scalable supervision for learning a machine-oriented quality metric without requiring human annotations or task-specific ground truth labels.

We further propose \emph{Multi-layer CLIP Similarity} (ML-CLIPSim), a differentiable full-reference metric learned from the proposed predictive-consistency supervision. CLIP visual representations~\cite{radford2021learning} provide transferable semantic features and have shown strong potential for image-quality-related tasks. However, using only the final global CLIP embedding can be too coarse for machine-oriented image quality assessment, especially when localized artifacts alter object boundaries, textures, or other discriminative regions important for downstream inference. ML-CLIPSim addresses this limitation by comparing reference and distorted images in a frozen CLIP feature space using two complementary sources of evidence: intermediate patch-token representations, which capture local and hierarchical visual changes, and global image embeddings, which encode semantic consistency. A lightweight learnable aggregation module combines these similarities, yielding a structured and differentiable proxy aligned with latent machine utility. The overall pipeline is shown in Fig.~\ref{fig:framework}.

Beyond quality assessment, ML-CLIPSim can be directly used as a distortion term for learned image compression. Modern learned codecs are commonly optimized with rate--distortion objectives based on MSE, MS-SSIM, or other perceptual losses~\cite{balle2018variational,jiang2025mlic}. These objectives, however, are not explicitly aligned with downstream machine performance. By replacing conventional distortion terms with ML-CLIPSim, a codec is encouraged to preserve visual information that is more predictive of machine inference, while remaining independent of any single downstream task model during deployment. This makes ML-CLIPSim suitable not only as an evaluation metric, but also as a reusable optimization objective for machine-oriented image coding.

The main contributions of this work are summarized as follows:
\begin{itemize}
    \item We formulate full-reference machine-oriented image quality as a latent machine utility and approximate it through predictive-consistency-based pairwise supervision across a population of downstream models.
    \item We introduce the Predictive Consistency Dataset for Machine Perception (PCMP), which contains PSNR-matched distortion pairs with soft preference labels generated by multi-model predictive-consistency voting. This design provides scalable supervision while reducing the influence of low-level distortion magnitude.
    \item We propose Multi-layer CLIP Similarity (ML-CLIPSim), a  differentiable quality metric built on frozen CLIP features. By aggregating intermediate patch-token similarities and global embedding similarities, ML-CLIPSim captures both localized evidence degradation and high-level semantic consistency.
    \item We demonstrate that ML-CLIPSim aligns better with machine-oriented preferences than conventional fidelity and perceptual metrics, generalizes to external machine-preference benchmarks, remains competitive on human-IQA datasets, and improves rate--task trade-offs when used as a distortion term for learned image compression.
\end{itemize}

% ------------------------------------------------------------
\section{Related Work}

\subsection{Perceptual Image Quality Assessment}
Image quality assessment (IQA) has been extensively studied for modeling perceptual image quality~\cite{zhai2020perceptual}. 
Early full-reference metrics such as SSIM and MS-SSIM~\cite{wang2004image,wang2003multiscale} improve upon pixel-wise fidelity by capturing structural similarity. 
Subsequent perceptual metrics, including FSIM and GMSD~\cite{zhang2011fsim,xue2013gradient}, further incorporate feature-based representations to better align with human visual perception. 
More recently, deep feature-based metrics such as LPIPS and DISTS~\cite{zhang2018unreasonable,ding2020image} significantly improve correlation with human judgments by leveraging pretrained neural network representations.

Pairwise preference learning (e.g., PieAPP, RankIQA~\cite{prashnani2018pieapp,liu2017rankiqa}) improves robustness for human perception but does not target machine-vision tasks.

\subsection{Machine-oriented Image Quality Assessment}
Recent work has begun to revisit IQA from the perspective of machine perception.
MPD~\cite{li2025image} shows that human-oriented IQA metrics often fail to reflect downstream model preferences, revealing a mismatch between perceptual quality and machine-oriented utility.

Different from prior work that mainly focuses on benchmark construction, we formulate machine-oriented quality as a latent utility and learn a differentiable surrogate through predictive consistency across multiple downstream models.

\subsection{CLIP-based Perceptual Modeling}
Foundation models such as CLIP~\cite{radford2021learning} provide transferable semantic representations and have recently been explored for image quality assessment.
Prior CLIP-based IQA methods mainly rely on global embeddings or prompt-based similarity~\cite{wang2023exploring,liao2026beyond,tang2024clip}.

In contrast, our method leverages multi-layer token representations under machine-oriented supervision, enabling the metric to better capture localized degradations and structured visual cues relevant to downstream machine perception.

\subsection{Machine-oriented Image Compression}
Coding for machines aims to optimize compression for downstream machine tasks rather than human perception~\cite{duan2020video,le2021image,Harell2023RateDistortionTI}.
Recent learned image compression methods achieve strong rate--distortion performance under human-oriented objectives such as MSE or MS-SSIM~\cite{balle2018variational,jiang2025mlic++}, but these objectives are often misaligned with downstream task performance.

Task-driven image compression methods address this issue by incorporating downstream task objectives or semantic representations into rate--distortion optimization~\cite{le2021image,Ge_2024_CVPR,yin2025unified}.
Among them, UG-ICM~\cite{yin2025unified} provides a unified task-agnostic framework based on CLIP supervision and serves as a strong baseline in our experiments.

Compared with prior approaches that are often task-specific or require task-dependent adaptation, our method learns a general-purpose machine-oriented quality metric from predictive consistency across multiple downstream models.

\section{Predictive Consistency Dataset for Machine Perception}
\label{sec:dataset}

\subsection{Latent Machine Utility and Pairwise Supervision}
We view machine-oriented image quality not as pixel fidelity, but as how well a distorted image preserves information for downstream inference. 
Formally, given a reference image $x$ and its distorted version $\hat{x}$, we define the latent machine utility as
\begin{equation}
U(x,\hat{x}) = \mathbb{E}_{m \sim \mathcal{P}(\mathcal{M})} \left[ u_m(x,\hat{x}) \right],
\end{equation}
where $\mathcal{P}(\mathcal{M})$ denotes a distribution over downstream models, and $u_m(x,\hat{x})$ measures prediction consistency under model $m$. 
In practice, we instantiate
\begin{equation}
u_m(x,\hat{x}) = - D_m\big(m(x), m(\hat{x})\big),
\end{equation}
where $D_m$ is a task-specific discrepancy function.

This formulation defines machine-oriented image quality as a latent functional over a population of downstream models, rather than as a fixed pixel-level distortion measure.

The utility $U(x,\hat{x})$ is latent because it depends on a large and diverse population of downstream models and tasks, which cannot be exhaustively observed or evaluated in practice. As a result, it cannot be reliably measured using any single downstream task or conventional distortion metric, making direct scalar supervision difficult.
This is because outputs from different models are not defined on a shared scale, making absolute quality scores difficult to calibrate consistently across tasks and architectures.

To address this limitation, we construct supervision through pairwise comparisons. Instead of assigning absolute quality scores, we compare distorted candidates relative to each other, since even models for the same task produce outputs that are not directly comparable (due to differences in scaling and prediction distributions).

Concretely, we build the Predictive Consistency Dataset for Machine Perception (PCMP) as an observable surrogate of the latent utility through pairwise supervision. For each pair, we aggregate preference votes from multiple pretrained models and use the resulting soft vote ratio as the supervision signal, reflecting the degree of agreement across a model population. To further reduce the influence of distortion magnitude, we restrict comparisons to PSNR-matched pairs, which conditions the supervision on similar distortion strength and emphasizes differences in task-relevant information preservation.

Each sample is formed from a reference image $x$ and a set of distorted variants $\{\hat{x}^{(p)}\}$ produced by controlled distortion processes. Unlike traditional IQA datasets, labels are derived from downstream-model \emph{predictive consistency} rather than human judgments.

\subsection{Distortion Generation}
We generate distorted variants using a diverse distortion library including traditional codecs (JPEG/WebP), pretrained learned codecs, resampling distortions, blur/noise perturbations, and color/tone transformations.
These distortions cover both model-sensitive and model-insensitive cases, enabling the dataset to capture distortions that may exhibit similar low-level fidelity while inducing different downstream-model behaviors.
Detailed distortion settings are provided in the supplementary material.

\subsection{PSNR-Matched Pair Sampling}
To control for low-level fidelity and isolate task-relevant effects, we form training pairs $(\hat{x}^{(p_0)},\hat{x}^{(p_1)})$ for the same reference image by requiring their PSNR values to be close:
\begin{equation}
\left|\mathrm{PSNR}(x,\hat{x}^{(p_0)}) - \mathrm{PSNR}(x,\hat{x}^{(p_1)})\right| \le \delta,
\end{equation}
where $\delta$ is a small tolerance (e.g., $0.5$ dB).
This PSNR-matching strategy reduces distortion-magnitude bias and yields harder, task-relevant pairs.

\subsection{Predictive Consistency and Label Assignment}

Let $\mathcal{M}=\{m_k\}_{k=1}^{K}$ denote a set of pretrained downstream models covering classification, detection, and segmentation. 
For each model $m_k$, we define a prediction discrepancy score between a reference image $x$ and its distorted version $\hat{x}$ as
\begin{equation}
d_k(x,\hat{x}) = D_k\!\left(m_k(x),\,m_k(\hat{x})\right),
\end{equation}
where $D_k$ is a task-specific discrepancy function.
Smaller values indicate better prediction consistency between $x$ and $\hat{x}$.

For classification, we use the KL divergence between predicted class distributions:
\begin{equation}
D^{\mathrm{cls}}_k
=
D_{\mathrm{KL}}\!\left(
\mathrm{softmax}(m_k(x))
\,\|\, 
\mathrm{softmax}(m_k(\hat{x}))
\right).
\end{equation}

For detection and instance segmentation, we use the reference prediction as pseudo ground truth.
Let $\{b_i,c_i\}_{i=1}^{N_x}$ and $\{\hat b_j,\hat c_j\}$ denote the predicted boxes and labels from $x$ and $\hat x$, respectively.
We perform label-aware matching:
\begin{equation}
j^*(i)=\arg\max_{j:\,\hat c_j=c_i}\mathrm{IoU}(b_i,\hat b_j).
\end{equation}
Let $w_i\in[0,1]$ denote the confidence score of the $i$-th reference prediction.

The detection discrepancy score is
\begin{equation}
D^{\mathrm{det}}_k(x,\hat x)
=
\frac{\sum_{i=1}^{N_x} w_i\, d_i^{\mathrm{det}}}{\sum_{i=1}^{N_x} w_i},
\end{equation}
where
\begin{equation}
d_i^{\mathrm{det}}=
\begin{cases}
1-\mathrm{IoU}(b_i,\hat b_{j^*(i)}), & \text{if a match exists},\\
1, & \text{otherwise}.
\end{cases}
\end{equation}

For instance segmentation and semantic segmentation, discrepancy scores are defined analogously using prediction consistency between the reference and distorted outputs.

Overall, $d_k$ measures how much a distortion perturbs the predictions of a given pretrained model.
A smaller $d_k$ indicates better preservation of task-relevant information without requiring task-specific supervision.

Given a PSNR-matched pair $(\hat{x}^{(p_0)},\hat{x}^{(p_1)})$, each model casts a vote:
\begin{equation}
v_k =
\begin{cases}
1, & d_k(x,\hat{x}^{(p_0)}) < d_k(x,\hat{x}^{(p_1)}) \\
0, & \text{otherwise}.
\end{cases}
\end{equation}
We define the continuous label as the vote ratio:
\begin{equation}
y\big(x,\hat{x}^{(p_0)},\hat{x}^{(p_1)}\big)
=
\frac{1}{K}\sum_{k=1}^{K} v_k
\in [0,1].
\end{equation}

The vote ratio reflects how consistently one distortion is preferred across diverse models, providing a robust soft supervision signal.
Compared with single-model labeling or hard binary supervision, multi-model voting with soft labels reduces model-specific bias and provides a more stable estimate of latent machine utility at the population level.
The vote ratio can be interpreted as an empirical estimate of the probability that one distortion is preferred under a randomly sampled downstream model.
% \begin{table}[t]
% \centering
% \scriptsize
% \setlength{\tabcolsep}{2pt}
% \renewcommand{\arraystretch}{0.9}
% \caption{Overview of the proposed Predictive Consistency Dataset (PCMP), including distortion generation, pair construction, and multi-model voting supervision.}
% \label{tab:dataset_overview}
% \scriptsize
% \setlength{\tabcolsep}{2pt}
% \renewcommand{\arraystretch}{1.15}
% \resizebox{0.7\columnwidth}{!}{%
% \begin{tabular}{l l}
% \toprule
% \textbf{Component} & \textbf{Setting} \\
% \midrule
% Reference set
% & 2000 images sampled from COCO train \\
% \midrule
% Distortion types
% & \makecell[l]{Codec: JPEG/WebP, learned codecs\\
% Resampling: down--up (bicubic/bilinear)\\
% Blur/Noise: Gaussian blur, noise\\
% Color/Tone: chroma and intensity shifts} \\
% \midrule
% Pair construction
% & \makecell[l]{PSNR-matched pairs ($\delta=0.5$ dB)\\
% to reduce low-level fidelity bias} \\
% \midrule
% Dataset size
% & \makecell[l]{66 variants per image; 327,241 pairs} \\
% \midrule
% Model pool
% & \makecell[l]{Classification: ResNet50, ViT-B/16, ConvNeXt-Tiny\\
% Detection: YOLOv8n, Faster R-CNN\\
% Segmentation: DeepLabV3, Mask R-CNN} \\
% \midrule
% Supervision
% & \makecell[l]{Pairwise preference label $y\in[0,1]$\\
% from multi-model consistency voting} \\
% \bottomrule
% \end{tabular}
% }
% \end{table}

\begin{figure}[t]
\centering
\includegraphics[width=.8\linewidth]{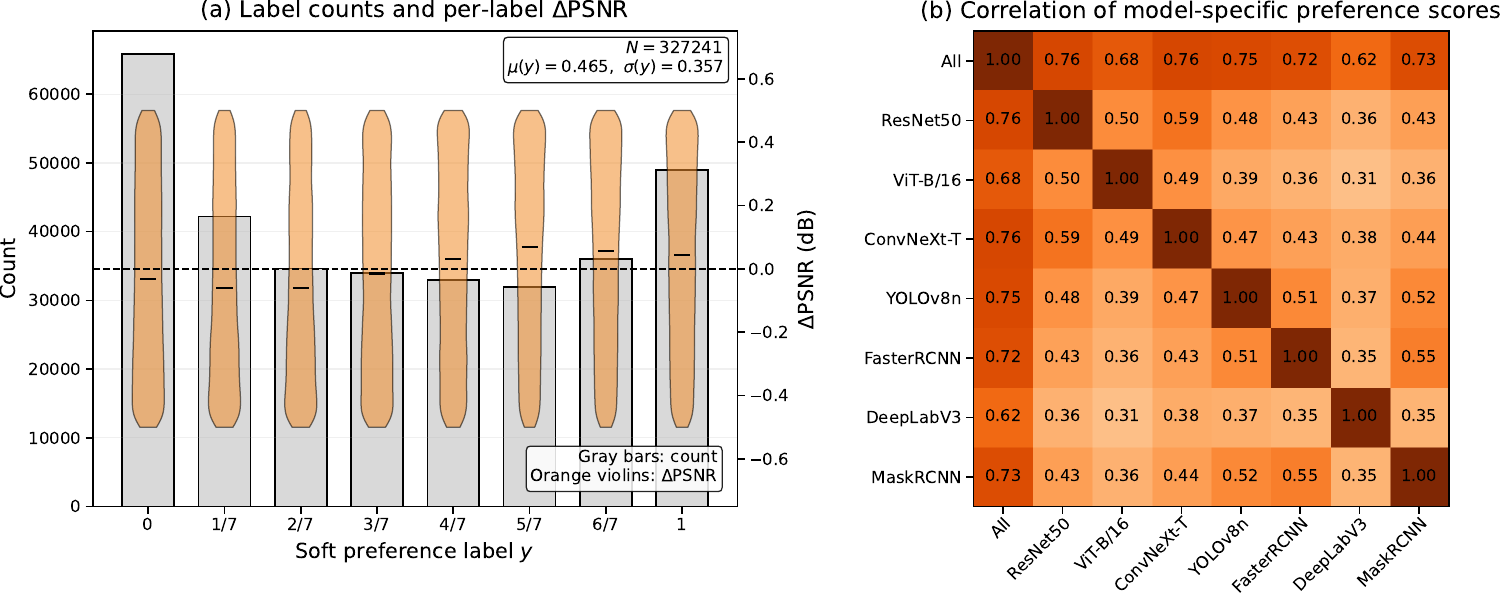}
\caption{Dataset statistics and model-level consistency of PCMP.
(a) Distribution of soft labels and PSNR differences.
(b) Correlation matrix of model-specific preference scores.}
\label{fig:dataset_stats_corr}
\end{figure}
\section{Method: Learning ML-CLIPSim}
\label{sec:method}

Given a clean reference image $x$ and its distorted or reconstructed version $\hat{x}$, our goal is to learn a differentiable scalar score
\begin{equation}
S_{\theta}(x,\hat{x}) \in \mathbb{R},
\end{equation}
where $\theta$ denotes the trainable parameters of the metric.

This score is trained to approximate the predictive consistency defined in Sec.~\ref{sec:dataset}, which serves as a proxy for latent machine utility. In particular, we aim to learn a differentiable function that aligns with the pairwise preference signals derived from multi-model voting and approximates the underlying utility through structured multi-layer feature aggregation.

A higher score indicates that $\hat{x}$ better preserves the information in $x$ that is useful to downstream machine-vision models.
The learned score is used both as an evaluation metric and as a proxy distortion term for learned image compression.
\subsection{Problem Formulation}
\label{sec:method_overview}
Unlike previous CLIP-based similarity methods~\cite{radford2021learning,wang2023exploring,liao2026beyond} that rely on final-layer global embeddings, comparing only the global representation of CLIP is often too coarse for compression. Localized degradations may significantly affect downstream inference while only weakly affecting the global embedding.
We adopt CLIP due to its strong semantic representation and transferability across diverse vision tasks.

From the perspective of latent machine utility, a suitable surrogate should capture both localized evidence corruption and global semantic consistency.
We therefore propose \emph{Multi-layer CLIP Similarity} (ML-CLIPSim), which combines multi-layer patch-token similarity with global CLIP embedding similarity.

\subsection{Multi-layer Token Similarity}
\label{sec:method_token}

We use the CLIP visual encoder (ViT-B/16) as a frozen backbone and extract features from transformer blocks, denoted by $\mathcal{L}$.
For each layer $l\in\mathcal{L}$, let
\begin{equation}
F_l(x)\in\mathbb{R}^{T\times D}
\end{equation}
denote the token feature of image $x$, where $D$ is the token dimension and $T$ is the total number of tokens.
Since CLIP ViT uses one CLS token and $P$ patch tokens, we have $T=P+1$.
The corresponding feature for the distorted image is denoted by $F_l(\hat{x})$.
All CLIP parameters are kept fixed during ML-CLIPSim training.

We directly L2-normalize the token features:
\begin{equation}
\tilde{F}_l(x)=\operatorname{L2norm}\!\left(F_l(x)\right),
\qquad
\tilde{F}_l(\hat{x})=\operatorname{L2norm}\!\left(F_l(\hat{x})\right),
\end{equation}
where $\operatorname{L2norm}(\cdot)$ denotes row-wise L2 normalization over the feature dimension $D$.

For each selected layer $l\in\mathcal{L}$, we compare the patch tokens of $x$ and $\hat{x}$.
Specifically, for patch-token index $t=1,\dots,P$ (the CLS token is index $0$), we compute
\begin{equation}
c_{l,t}(x,\hat{x})
=
\left\langle
\tilde{F}_{l,t}(x),\,
\tilde{F}_{l,t}(\hat{x})
\right\rangle .
\label{eq:token_cosine}
\end{equation}
Here $\tilde{F}_{l,t}(x)\in\mathbb{R}^{D}$ is the $t$-th normalized patch token from layer $l$, and $c_{l,t}(x,\hat{x})\in[-1,1]$ is the corresponding token similarity.

We first summarize patch similarity at layer $l$ by mean pooling:
\begin{equation}
s_l(x,\hat{x})
=
\frac{1}{P}\sum_{t=1}^{P} c_{l,t}(x,\hat{x}).
\label{eq:uniform_pool}
\end{equation}

Specifically, we partition the 12 CLIP layers in $\mathcal{L}$ into $G=3$ contiguous groups, corresponding to early, middle, and late stages.
We assign a learnable weight to each group, while distributing the weight uniformly among the layers within the same group.

Let $\mathbf{w}\in\mathbb{R}^{G}$ denote learnable group logits, normalized via softmax:
\begin{equation}
\pi_g
=
\frac{\exp(w_g)}
{\sum_{j=1}^{G}\exp(w_j)},
\qquad g=1,\dots,G,
\end{equation}
such that $\pi_g \ge 0$ and $\sum_{g=1}^{G}\pi_g=1$.

Let $\mathcal{L}_g\subseteq\mathcal{L}$ denote the layers in group $g$, we assign the group weight uniformly to its layers:
\begin{equation}
\alpha_l = \frac{\pi_g}{|\mathcal{L}_g|}, \qquad l \in \mathcal{L}_g.
\end{equation}

By construction, the resulting layer weights satisfy $\sum_{l\in\mathcal{L}} \alpha_l = 1$.
The token-branch score is then computed as
\begin{equation}
S_{\text{token}}(x,\hat{x})
=
\sum_{l\in\mathcal{L}} \alpha_l\, s_l(x,\hat{x}),
\end{equation}
where $s_l(x,\hat{x})$ is the similarity score from layer $l$.
\begin{table}[t]
\centering
\caption{Performance comparison on the PCMP test set. Higher is better.}
\label{tab:metric_main}
\resizebox{0.65\columnwidth}{!}{%
\begin{tabular}{l c c c c}
\toprule
\textbf{Metric} & \textbf{ACC}$\uparrow$&\textbf{SRCC}$\uparrow$&\textbf{KRCC}$\uparrow$&\textbf{PLCC}$\uparrow$\\
\midrule
PSNR
& 0.6196 & 0.0291 & 0.0622 & 0.1280 \\
MS-SSIM
& 0.7301 & 0.4961 & 0.3745 & 0.3711 \\
DISTS
& 0.7686 & 0.6663 & 0.4933 & 0.6317 \\
LPIPS
& 0.8042 & 0.7144 & 0.5374 & 0.6513 \\
CLIP Score
& 0.8030 & 0.6953 & 0.5150 & 0.6099 \\
% \textbf{ML-CLIPSim (Ours)}
% & \textbf{0.8063} & \textbf{0.7830} & \textbf{0.6040} & \textbf{0.7317} \\
% \midrule
\textbf{ML-CLIPSim (Ours)}
& \textbf{0.8284} & \textbf{0.7937} & \textbf{0.6182} & \textbf{0.7364} \\

% Metric          ACC       PLCC       SRCC       KRCC
% ----------------------------------------------------
% ours         0.8284     0.7364     0.7937     0.6182
% psnr         0.6173     0.1464     0.0416     0.0729
% msssim       0.7315     0.3799     0.5056     0.3859
% dists        0.7633     0.6225     0.6592     0.4897
% lpips        0.8024     0.6505     0.7156     0.5420
\bottomrule
\end{tabular}
}
% \vspace{-6pt}
\end{table}

\subsection{Global Similarity and Final ML-CLIPSim Score}
\label{sec:method_global}

To complement token-level matching, we also compare the image-level CLIP embedding.
Let $g(x), g(\hat{x})\in\mathbb{R}^{D_g}$ denote the normalized global CLIP image embeddings of the reference and distorted images after the standard CLIP visual projection.
Their cosine similarity is
\begin{equation}
S_{\text{global}}(x,\hat{x})
=
\left\langle
g(x),\,
g(\hat{x})
\right\rangle.
\label{eq:global_branch}
\end{equation}
Since both embeddings are L2-normalized, $S_{\text{global}}(x,\hat{x})\in[-1,1]$.

We combine $S_{\text{token}}$ and $S_{\text{global}}$ through a learned scalar gate
\begin{equation}
\eta=\sigma(w_{\text{gate}})\in(0,1),
\label{eq:gate}
\end{equation}
where $w_{\text{gate}}\in\mathbb{R}$ is a learnable scalar and $\sigma(\cdot)$ is the sigmoid function.
The final ML-CLIPSim score is defined as
\begin{equation}
S_{\theta}(x,\hat{x})
=
\eta\, S_{\text{token}}(x,\hat{x})
+
(1-\eta)\, S_{\text{global}}(x,\hat{x}).
\label{eq:ML-CLIPSim_final}
\end{equation}

\subsection{Training with Continuous Pairwise Preferences}
\label{sec:metric_training}
We train ML-CLIPSim using the PSNR-matched pairwise dataset in Sec.~\ref{sec:dataset}. Each example consists of a reference image $x$, two distorted versions $\hat{x}^{(p_0)}$ and $\hat{x}^{(p_1)}$, and a continuous target
\begin{equation}
y\big(x,\hat{x}^{(p_0)},\hat{x}^{(p_1)}\big)\in[0,1],
\end{equation}
where $p_0$ and $p_1$ are the indices of the two distortion candidates and $y\big(x,\hat{x}^{(p_0)},\hat{x}^{(p_1)}\big)$ is the vote ratio of downstream models preferring $\hat{x}^{(p_0)}$ over $\hat{x}^{(p_1)}$.
Following Sec.~\ref{sec:dataset}, we discard ambiguous ties with $y=0.5$.

We model pairwise preference through the score difference
\begin{equation}
z_{\theta}
=
S_{\theta}(x,\hat{x}^{(p_0)})
-
S_{\theta}(x,\hat{x}^{(p_1)}).
\label{eq:pair_logit}
\end{equation}
This formulation assumes that pairwise preference probabilities can be approximated by a logistic function of latent utility differences. We therefore train ML-CLIPSim with BCE on the pairwise logit and the soft vote-ratio label:
\begin{equation}
\mathcal{L}_{\text{metric}}
=
\operatorname{BCEWithLogits}\Big(z_\theta,\, y\big(x,\hat{x}^{(p_0)},\hat{x}^{(p_1)}\big)\Big),
\label{eq:metric_loss}
\end{equation}

\subsection{Using ML-CLIPSim for Codec Optimization}
\label{sec:codec_opt}

After training, ML-CLIPSim can be used as a differentiable proxy distortion term for learned image compression.
Given an input image $x$, a codec encodes $x$ into a bitstream and reconstructs $\hat{x}$, yielding a rate term $R$.
We optimize the following rate--distortion objective:
\begin{equation}
\mathcal{L}_{\text{codec}}
=
R + \lambda D_{\mathrm{ML-CLIPSim}},
\quad
D_{\mathrm{ML-CLIPSim}} = 1 - S_{\theta}(x,\hat{x}),
\label{eq:codec_loss}
\end{equation}
where $\lambda>0$ is the trade-off parameter between bitrate and machine-oriented distortion.

Minimizing Eq.~\eqref{eq:codec_loss} encourages the codec to preserve the visual content that is most predictive of downstream machine tasks. From the latent utility perspective, this objective can be interpreted as a surrogate-based rate--utility optimization, where ML-CLIPSim provides a differentiable approximation of machine-oriented utility. During codec training, the ML-CLIPSim network is kept fixed.

\begin{table}[t]
\centering
\small
\setlength{\tabcolsep}{3pt}
\renewcommand{\arraystretch}{1.1}
\caption{Correlation with preference labels on MPD. Best and second-best results are highlighted in red and blue, respectively.}
\label{tab:mpd_main}
\resizebox{\linewidth}{!}{
\begin{tabular}{l|ccc|ccc|ccc|ccc|ccc}
\toprule
\multirow{2}{*}{Metric}
& \multicolumn{3}{c|}{Severe Distortion}
& \multicolumn{3}{c|}{Mild Distortion}
& \multicolumn{3}{c|}{NSI}
& \multicolumn{3}{c|}{SCI}
& \multicolumn{3}{c}{AIGI} \\
% \cmidrule(lr){2-4}\cmidrule(lr){5-7}\cmidrule(lr){8-10}\cmidrule(lr){11-13}\cmidrule(lr){14-16}
& SRCC$\uparrow$ & KRCC$\uparrow$ & PLCC$\uparrow$
& SRCC$\uparrow$ & KRCC$\uparrow$ & PLCC$\uparrow$
& SRCC$\uparrow$ & KRCC$\uparrow$ & PLCC$\uparrow$
& SRCC$\uparrow$ & KRCC$\uparrow$ & PLCC$\uparrow$
& SRCC$\uparrow$ & KRCC$\uparrow$ & PLCC$\uparrow$ \\
\midrule
PSNR
& 0.176 & 0.118 & 0.262
& 0.248 & 0.170 & 0.484
& 0.345 & 0.239 & 0.452
& 0.310 & 0.209 & 0.406
& 0.339 & 0.234 & 0.452 \\
MS-SSIM
& 0.481 & 0.329 & 0.486
& 0.395 & 0.270 & 0.436
& 0.571 & 0.399 & 0.569
& 0.529 & 0.367 & 0.538
& 0.510 & 0.353 & 0.517 \\
LPIPS
& 0.669 & 0.479 & 0.678
& 0.626 & 0.444 & 0.625
& 0.736 & 0.543 & 0.730
& 0.720 & 0.527 & 0.727
& 0.660 & 0.474 & 0.681 \\
DISTS
& \textcolor{blue}{0.770} & \textcolor{blue}{0.571} & \textcolor{blue}{0.778}
& \textcolor{blue}{0.731} & \textcolor{blue}{0.538} & \textcolor{blue}{0.741}
& \textcolor{blue}{0.816} & \textcolor{blue}{0.622} & \textcolor{blue}{0.823}
& \textcolor{blue}{0.788} & \textcolor{blue}{0.593} & \textcolor{blue}{0.799}
& \textcolor{red}{0.764} & \textcolor{red}{0.569} & \textcolor{red}{0.777} \\

CLIPScore
& 0.735 & 0.540 & 0.752
& 0.716 & 0.523 & 0.704
& 0.806 & 0.614 & 0.802
& 0.779 & 0.585 & 0.784
& 0.702 & 0.508 & 0.708 \\

% ML-CLIPSim (Ours)
% & \textcolor{red}{0.783} & \textcolor{red}{0.588} & \textcolor{red}{0.798}
% & \textcolor{red}{0.741} & \textcolor{red}{0.550} & \textcolor{red}{0.748}
% & \textcolor{red}{0.827} & \textcolor{red}{0.637} & \textcolor{red}{0.833}
% & \textcolor{red}{0.828} & \textcolor{red}{0.634} & \textcolor{red}{0.828}
% & \textcolor{blue}{0.751} & \textcolor{blue}{0.554} & \textcolor{blue}{0.762} \\
% ML-CLIPSim (Ours)
% & 0.785 & 0.590 & 0.801 & 0.743 & 0.551 & 0.749
% & 0.828 & 0.638 & 0.835
% & 0.832 & 0.638 & 0.831
% & 0.750 & 0.554 & 0.762 \\
\textbf{ML-CLIPSim (Ours)}
& \textbf{\textcolor{red}{0.783}} & \textbf{\textcolor{red}{0.588}} & \textbf{\textcolor{red}{0.799}}
& \textbf{\textcolor{red}{0.741}} & \textbf{\textcolor{red}{0.549}} & \textbf{\textcolor{red}{0.747}}
& \textbf{\textcolor{red}{0.826}} & \textbf{\textcolor{red}{0.635}} & \textbf{\textcolor{red}{0.832}}
& \textbf{\textcolor{red}{0.833}} & \textbf{\textcolor{red}{0.640}} & \textbf{\textcolor{red}{0.832}}
& \textbf{\textcolor{blue}{0.748}} & \textbf{\textcolor{blue}{0.552}} & \textbf{\textcolor{blue}{0.760}} \\
\midrule
\multirow{2}{*}{Metric}
& \multicolumn{3}{c|}{YoN}
& \multicolumn{3}{c|}{MCQ}
& \multicolumn{3}{c|}{VQA}
& \multicolumn{3}{c|}{CAP}
& \multicolumn{3}{c}{Others} \\
% \cmidrule(lr){2-4}\cmidrule(lr){5-7}\cmidrule(lr){8-10}\cmidrule(lr){11-13}\cmidrule(lr){14-16}
& SRCC$\uparrow$ & KRCC$\uparrow$ & PLCC$\uparrow$
& SRCC$\uparrow$ & KRCC$\uparrow$ & PLCC$\uparrow$
& SRCC$\uparrow$ & KRCC$\uparrow$ & PLCC$\uparrow$
& SRCC$\uparrow$ & KRCC$\uparrow$ & PLCC$\uparrow$
& SRCC$\uparrow$ & KRCC$\uparrow$ & PLCC$\uparrow$ \\
\midrule
PSNR
& 0.267 & 0.182 & 0.344
& 0.341 & 0.234 & 0.438
& 0.208 & 0.139 & 0.239
& 0.376 & 0.259 & 0.457
& 0.208 & 0.140 & 0.323 \\
MS-SSIM
& 0.409 & 0.280 & 0.424
& 0.539 & 0.374 & 0.529
& 0.298 & 0.201 & 0.299
& 0.587 & 0.412 & 0.590
& 0.390 & 0.264 & 0.391 \\
LPIPS
& 0.507 & 0.353 & 0.538
& 0.693 & 0.501 & 0.685
& 0.416 & 0.284 & 0.430
& 0.731 & 0.537 & 0.741
& 0.511 & 0.354 & 0.519 \\
DISTS
& \textcolor{blue}{0.563} & \textcolor{blue}{0.396} & \textcolor{blue}{0.595}
& \textcolor{blue}{0.753} & 0.557 & \textcolor{blue}{0.756}
& 0.445 & 0.306 & \textcolor{blue}{0.472}
& \textcolor{blue}{0.780} & \textcolor{blue}{0.588} & \textcolor{blue}{0.797}
& \textcolor{blue}{0.613} & \textcolor{blue}{0.433} & \textcolor{blue}{0.626} \\
CLIPScore
& 0.556 & 0.393 & 0.586
& 0.759 & \textcolor{blue}{0.565} & 0.753
& \textcolor{red}{0.451} & \textcolor{red}{0.309} & \textcolor{red}{0.478}
& 0.763 & 0.570 & 0.768
& 0.540 & 0.379 & 0.545 \\

% ML-CLIPSim (Ours)
% & \textcolor{red}{0.569} & \textcolor{red}{0.402} & \textcolor{red}{0.608}
% & \textcolor{red}{0.776} & \textcolor{red}{0.582} & \textcolor{red}{0.779}
% & \textcolor{red}{0.452} & \textcolor{red}{0.309} & 0.471
% & \textcolor{red}{0.792} & \textcolor{red}{0.599} & \textcolor{red}{0.811}
% & \textcolor{red}{0.619} & \textcolor{red}{0.439} & \textcolor{blue}{0.626} \\
% ML-CLIPSim (Ours)
% & 0.570 & 0.403 & 0.609
% & 0.778 & 0.584 & 0.780
% & 0.451 & 0.308 & 0.471
% & 0.794 & 0.601 & 0.813
% & 0.621 & 0.441 & 0.628 \\
\textbf{ML-CLIPSim (Ours)}
& \textbf{\textcolor{red}{0.569}} & \textbf{\textcolor{red}{0.402}} & \textbf{\textcolor{red}{0.607}}
& \textbf{\textcolor{red}{0.775}} & \textbf{\textcolor{red}{0.581}} & \textbf{\textcolor{red}{0.777}}
& \textbf{\textcolor{blue}{0.449}} & \textbf{\textcolor{blue}{0.306}} & \textbf{0.468}
& \textbf{\textcolor{red}{0.792}} & \textbf{\textcolor{red}{0.599}} & \textbf{\textcolor{red}{0.811}}
& \textbf{\textcolor{red}{0.623}} & \textbf{\textcolor{red}{0.442}} & \textbf{\textcolor{red}{0.630}} \\
\bottomrule
\end{tabular}%
}
% \vspace{-6pt}
\end{table}

\section{Experiments}
\begin{figure*}[t]
\centering
\setlength{\tabcolsep}{0pt}

\begin{tabular}{@{}c@{\hspace{0.02\textwidth}}c@{}}
\includegraphics[width=.45\textwidth]{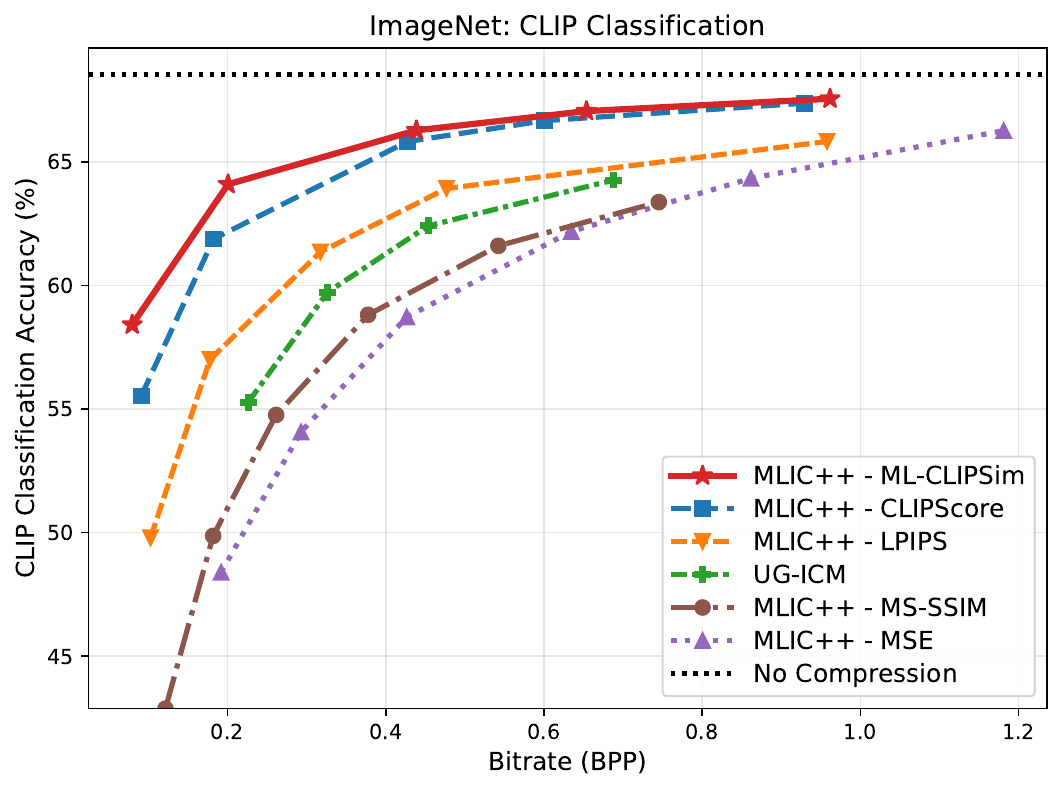} &
\includegraphics[width=.45\textwidth]{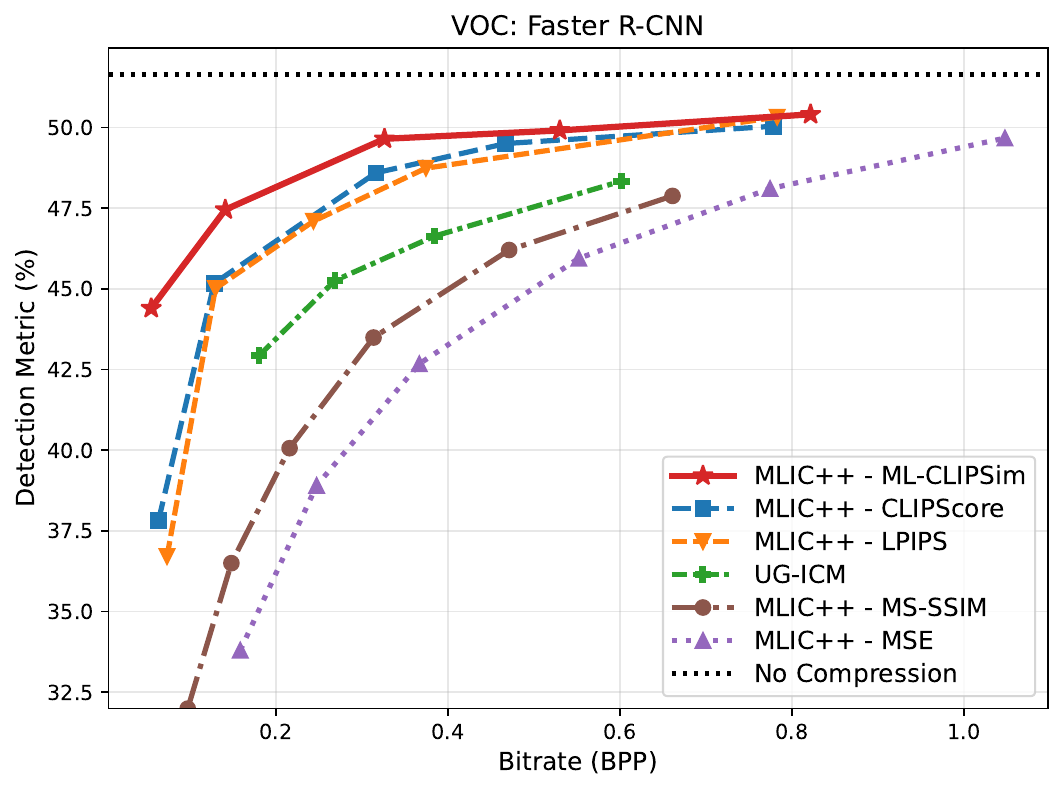} \\[-2pt]
\includegraphics[width=.45\textwidth]{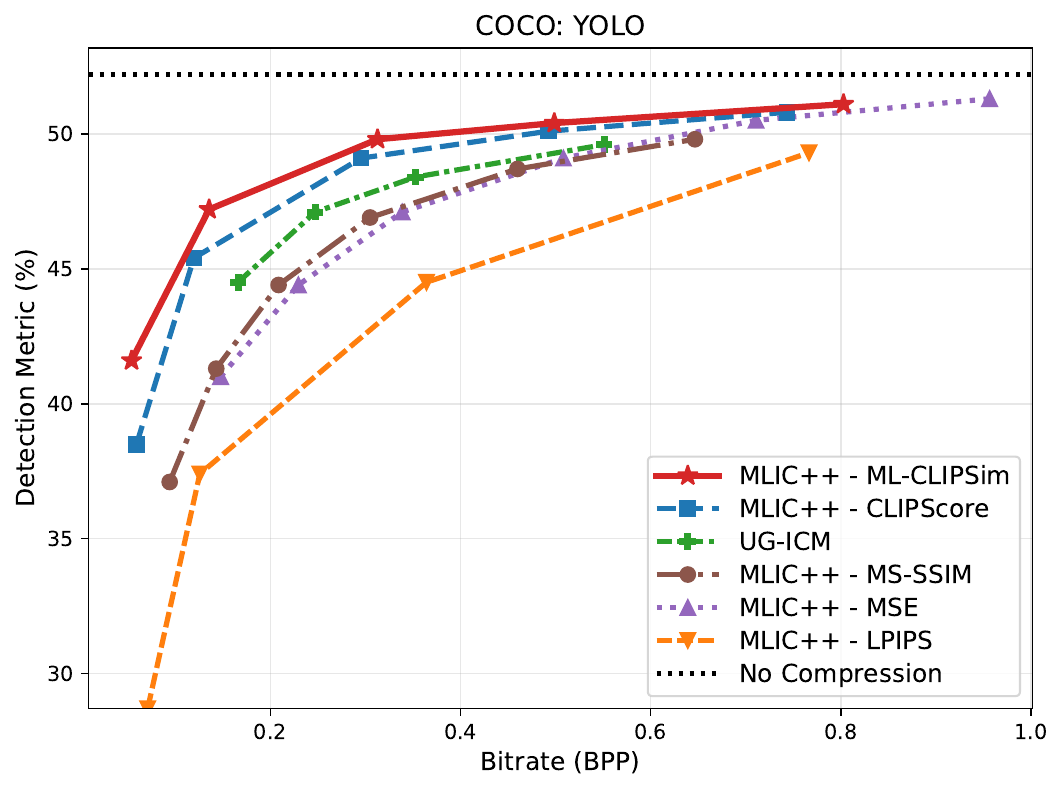} &
\includegraphics[width=.45\textwidth]{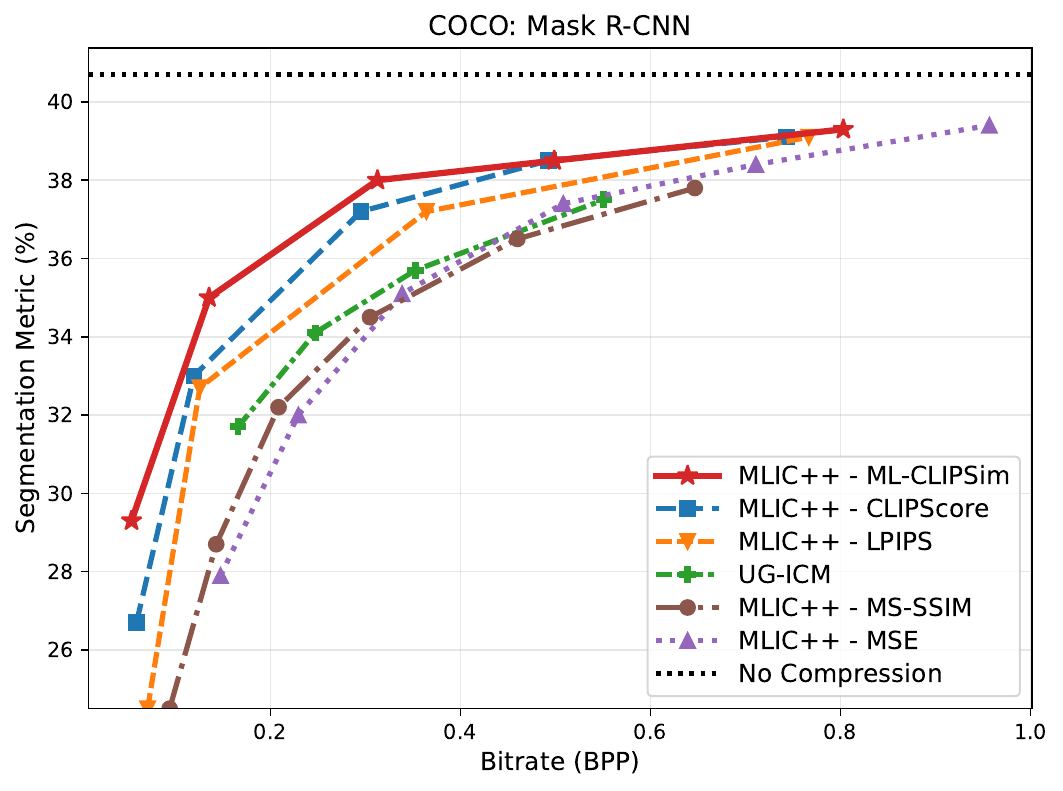} \\[-2pt]
\includegraphics[width=.45\textwidth]{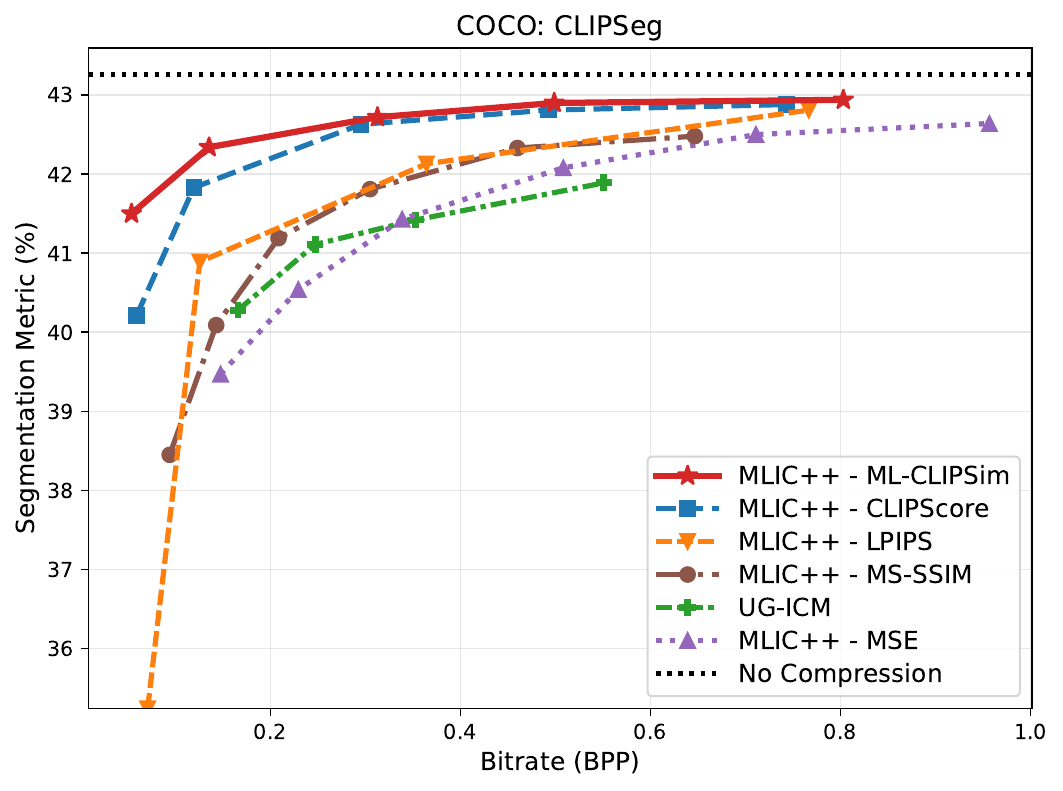} &
\includegraphics[width=.45\textwidth]{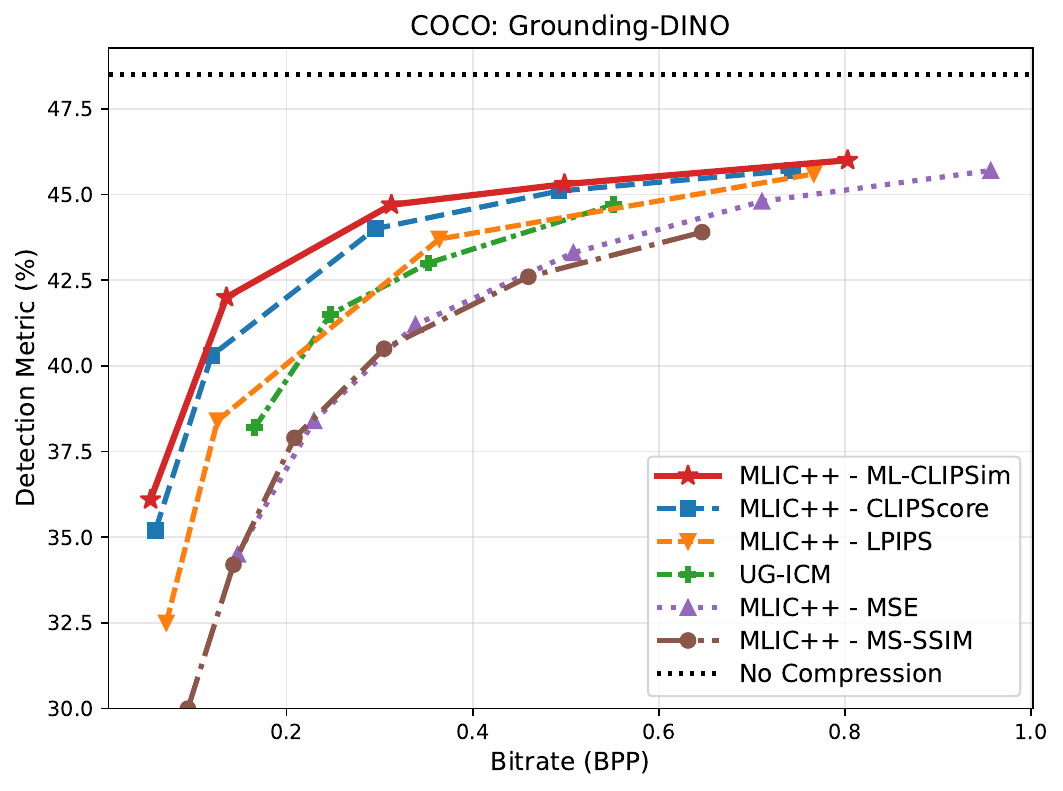}
\end{tabular}

\caption{Rate--task curves across ImageNet, VOC, and COCO downstream tasks. ML-CLIPSim consistently achieves the best trade-off across tasks.}
\label{fig:codec_all_curves}
\end{figure*}
\label{sec:experiments}

\noindent\textbf{Overview.}
We evaluate the proposed pipeline in three steps: (i) dataset construction and label distribution, (ii) learning the ML-CLIPSim metric, and (iii) using ML-CLIPSim as the distortion term for training learned image codecs.

% ------------------------------------------------------------
\subsection{Dataset Construction}
We randomly sample 2000 images from the COCO training split as references~\cite{lin2014microsoft}.
For each reference image, we generate 66 distorted variants using the distortion library described in Sec.~\ref{sec:dataset}, including traditional codecs (JPEG/WebP), pretrained learned codecs, resampling distortions, blur/noise perturbations, and color/tone transformations.
For label generation, we use a voter pool spanning classification, detection, and segmentation models, including ResNet50, ViT-B/16, ConvNeXt-Tiny, YOLOv8n, Faster R-CNN, DeepLabV3, and Mask R-CNN.
We form training pairs $(\hat{x}^{(p_0)},\hat{x}^{(p_1)})$ only when their PSNR values are within $\delta=0.5$~dB, and assign a continuous preference label $y\in[0,1]$ from multi-model predictive-consistency voting.
The final dataset contains 327,241 PSNR-matched pairs.

We randomly split the labeled pairs into an 8:2 train/test partition and evaluate generalization on MPD~\cite{li2025image}.
Fig.~\ref{fig:dataset_stats_corr} illustrates the label distribution and model-level consistency of the constructed dataset.
The labels are well distributed across different vote-ratio levels, and substantial positive correlations are observed across different downstream models, suggesting that the collected supervision captures shared machine-oriented preference patterns.

% \begin{figure}[t]
% \centering
% \setlength{\tabcolsep}{0pt}
% \begin{tabular}{@{}c@{\hspace{0.02\columnwidth}}c@{}}
% \includegraphics[width=.45\linewidth]{paper_figures_combined_scheme1/imagenet_clipcls_combined_scheme1.pdf} &
% \includegraphics[width=.45\linewidth]{paper_figures_combined_scheme1/voc_frcnn_combined_scheme1.pdf} \\
% \end{tabular}
% \caption{Rate--task curves on ImageNet classification and VOC detection. ML-CLIPSim achieves the best trade-off across both tasks.}
% \label{fig:codec_all_curves_a}
% \end{figure}

% \begin{figure}[t]
% \centering
% \setlength{\tabcolsep}{0pt}
% \begin{tabular}{@{}c@{\hspace{0.02\columnwidth}}c@{}}
% \includegraphics[width=.45\linewidth]{paper_figures_combined_scheme1/coco_yolo_combined_scheme1.pdf} &
% \includegraphics[width=.45\linewidth]{paper_figures_combined_scheme1/coco_maskrcnn_combined_scheme1.pdf} \\
% \includegraphics[width=.45\linewidth]{paper_figures_combined_scheme1/coco_clipseg_combined_scheme1.pdf} &
% \includegraphics[width=.45\linewidth]{paper_figures_combined_scheme1/coco_groundingdino_combined_scheme1.pdf} \\
% \end{tabular}
% \caption{Rate--task curves on COCO downstream tasks. ML-CLIPSim consistently achieves the best trade-off across tasks.}
% \label{fig:codec_all_curves_c}
% \end{figure}

\begin{figure}[t]
\centering
\setlength{\tabcolsep}{0pt}
\begin{tabular}{@{}c@{\hspace{0.01\columnwidth}}c@{}}
\includegraphics[width=.45\linewidth]
{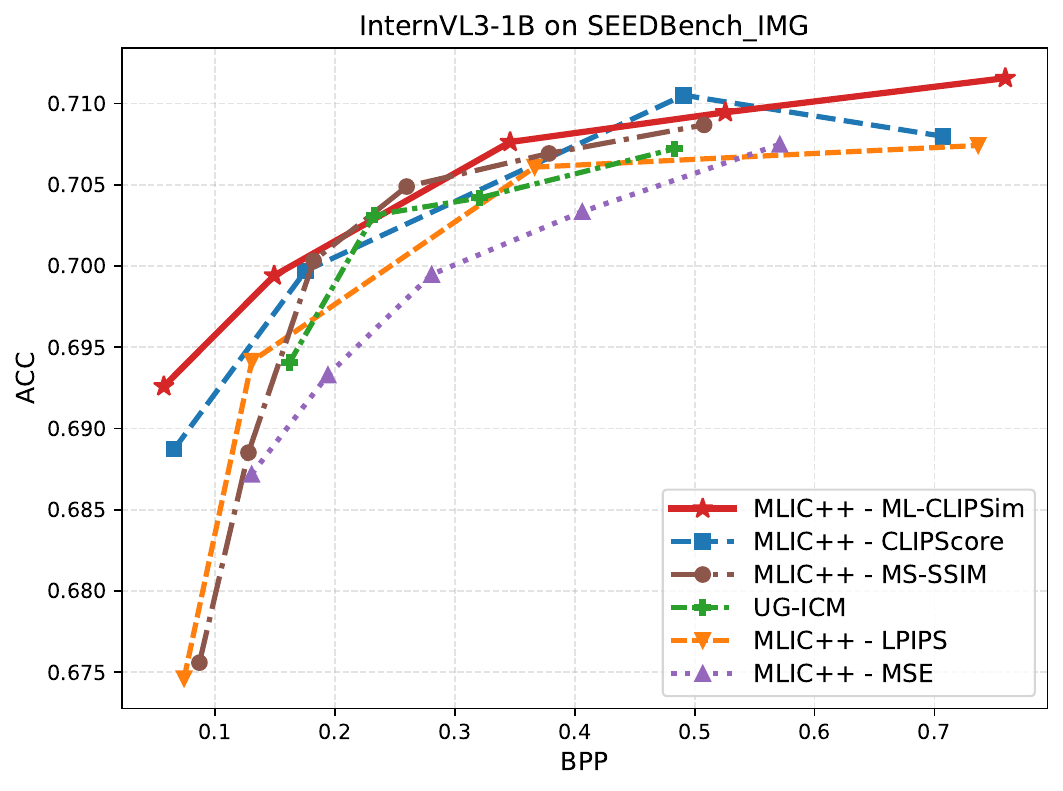} &
\includegraphics[width=.45\linewidth]{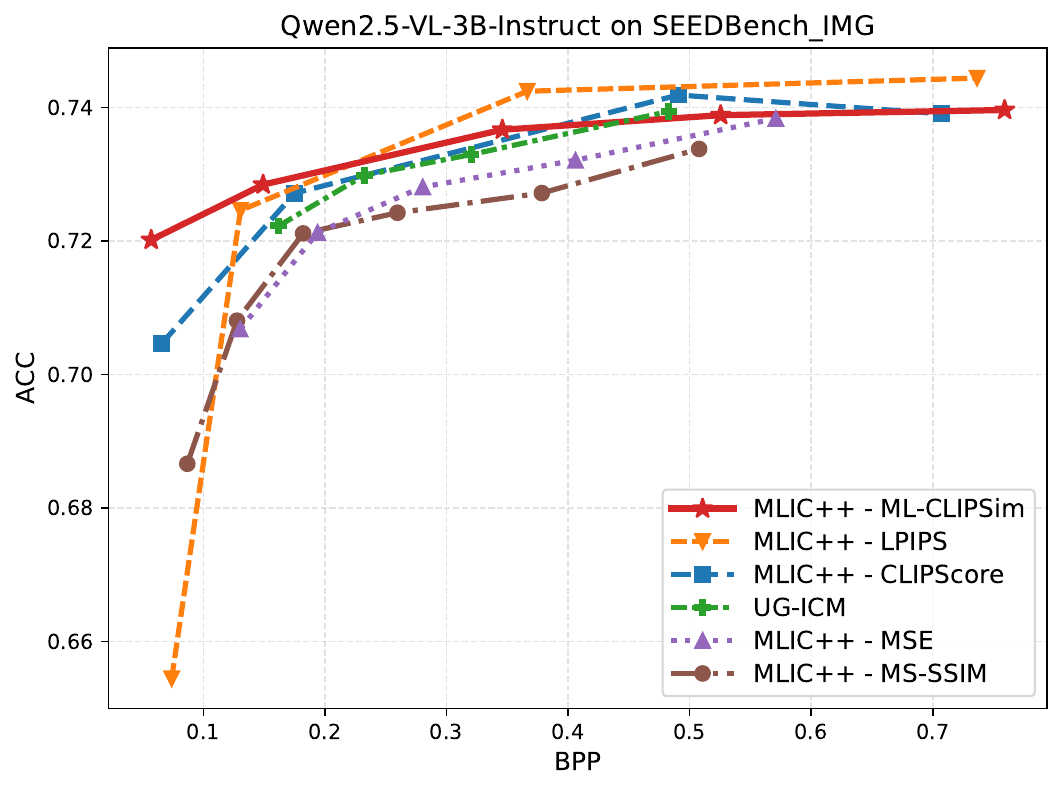} \\

\includegraphics[width=.45\linewidth]{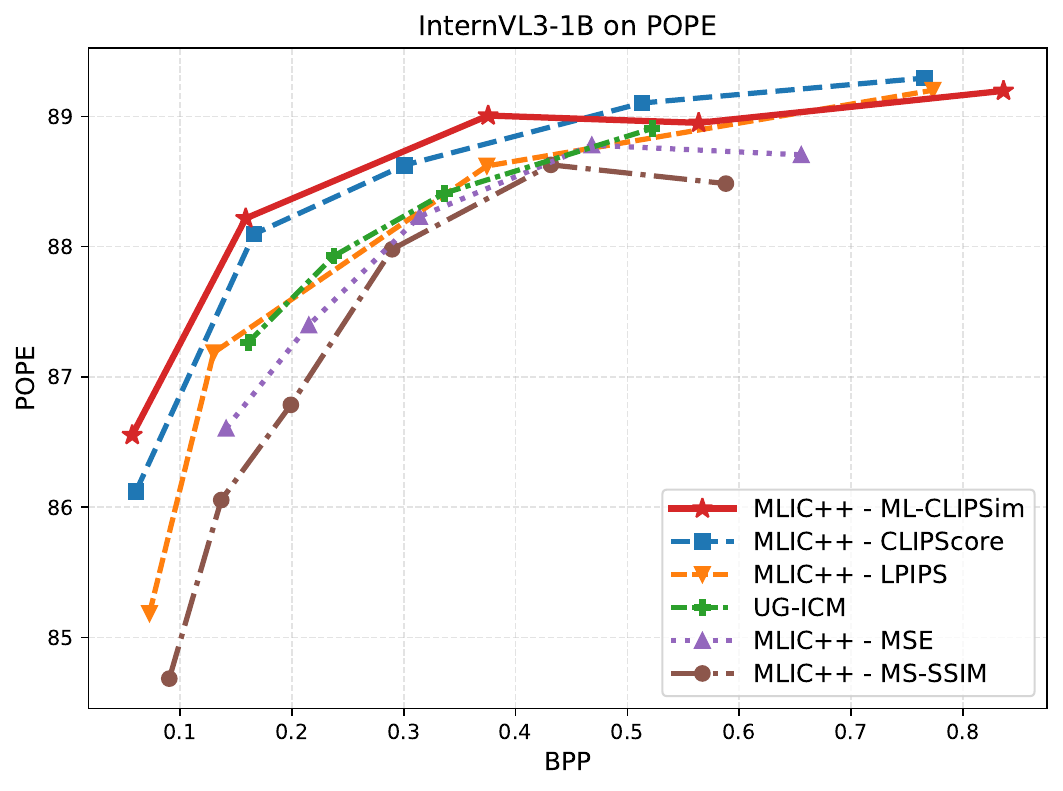} &
\includegraphics[width=.45\linewidth]{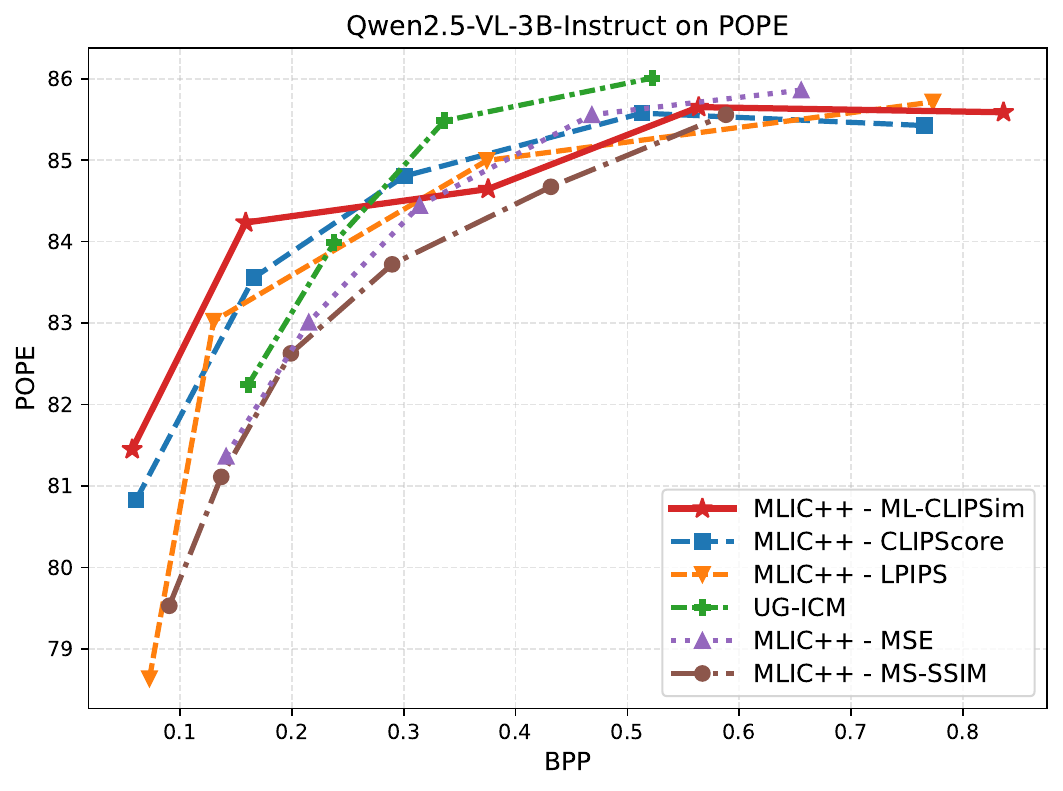} \\
\end{tabular}
\caption{VLM evaluation on SEEDBench and POPE using InternVL3-1B and Qwen2.5-VL-3B-Instruct. ML-CLIPSim achieves the best rate--task performance across benchmarks.}
\label{fig:vlm_main}
\end{figure}

% ------------------------------------------------------------
\subsection{Training and Evaluating ML-CLIPSim}
We train ML-CLIPSim on the PCMP pairs using the objective defined in Sec.~\ref{sec:metric_training}. 
Specifically, we use the Adam optimizer with a learning rate of $2\times10^{-4}$ and a batch size of 64, and train for 5 epochs with an 80/20 train/validation split (random seed 123). 
Only the lightweight aggregation parameters are updated while the CLIP backbone remains frozen.

Table~\ref{tab:metric_main} compares ML-CLIPSim against common baselines, including PSNR, MS-SSIM, DISTS, LPIPS, and global CLIP similarity (CLIPScore), on the PCMP test set. ML-CLIPSim consistently achieves the best performance across metrics, demonstrating its effectiveness in capturing machine-oriented perceptual differences beyond traditional and learned baselines.

\noindent\textbf{Cross-Distortion and Cross-Content Generalization.}
To evaluate the generalization ability of ML-CLIPSim beyond PCMP, we further test it on an external machine-vision quality assessment benchmark, the Machine Preference Database (MPD)~\cite{li2025image}, which differs significantly from our data in both distortion types and label construction protocol.

We evaluate on the public MPD split, which includes diverse distortion categories and downstream task settings.
Table~\ref{tab:mpd_main} shows that ML-CLIPSim generalizes well and outperforms CLIPScore, LPIPS, and DISTS on most MPD subsets.

\noindent\textbf{Human-IQA Benchmark Results.}
Although ML-CLIPSim is trained with machine-perception supervision, we also evaluate its correlation with human opinion scores on standard IQA datasets (TID2013~\cite{ponomarenko2015image} and LIVE~\cite{sheikh2006statistical}).
Following common practice, we report PLCC, SRCC, and KRCC for human-IQA evaluation, and compare with classical and learned perceptual metrics including MS-SSIM, DISTS and LPIPS.
This comparison indicates partial consistency between machine- and human-perception characteristics. Although ML-CLIPSim is optimized for machine-vision objectives, it remains competitive on human-perception metrics.
Results are summarized in Table~\ref{tab:iqa_main}.

\begin{table}[]
\centering
\small
\setlength{\tabcolsep}{3pt}
\renewcommand{\arraystretch}{0.9}
\caption{Correlation on TID2013 and LIVE human-IQA benchmarks. Higher is better.}
\label{tab:iqa_main}
\resizebox{0.75\columnwidth}{!}{%
\begin{tabular}{l|ccc|ccc}
\toprule
\multirow{2}{*}{\textbf{Metric}} 
& \multicolumn{3}{c|}{\textbf{TID2013}} 
& \multicolumn{3}{c}{\textbf{LIVE}} \\
\cmidrule(lr){2-4}\cmidrule(lr){5-7}
&
\textbf{SRCC}$\uparrow$&\textbf{KRCC}$\uparrow$&\textbf{PLCC}$\uparrow$&
\textbf{SRCC}$\uparrow$&\textbf{KRCC}$\uparrow$&\textbf{PLCC}$\uparrow$\\
\midrule

PSNR
& 0.687 & 0.496 & 0.677
& 0.873 & 0.680 & 0.865 \\

MS-SSIM
& 0.786 & 0.605 & 0.830
& {0.951} & {0.803} & 0.940 \\

LPIPS
& 0.670 & 0.497 & 0.749
& 0.924 & 0.751 & 0.914 \\

DISTS
& {0.830} & {0.639} & {0.855}
& 0.948 & 0.793 & {0.944} \\

CLIPScore
& 0.744 & 0.552 & 0.794
& 0.892 & 0.712 & 0.903 \\
% \midrule
\textbf{ML-CLIPSim (ours)}
& {0.734} & {0.551} & {0.798}
& {0.929} & {0.760} & {0.934} \\

\bottomrule
\end{tabular}%
}
\end{table}

% ------------------------------------------------------------
% ------------------------------------------------------------
\subsection{Learned Image Compression Results}
\begin{table}[t]
\centering
\caption{BD-Rate (\%) comparison across datasets using MLIC++-MSE as the baseline. Lower is better. Best and second-best results are highlighted in red and blue, respectively.}
\label{tab:bd_rate_dataset}
\resizebox{\linewidth}{!}{
\begin{tabular}{l|c|cccc|c|cc|cc}
\toprule
\multirow{2}{*}{\textbf{Method}}
& \multicolumn{1}{c|}{\textbf{ImageNet}}
& \multicolumn{4}{c|}{\textbf{COCO}}
& \multicolumn{1}{c|}{\textbf{VOC}}
& \multicolumn{2}{c|}{\textbf{VLM POPE}}
& \multicolumn{2}{c}{\textbf{VLM SeedBench}} \\
\cmidrule(lr){2-2} \cmidrule(lr){3-6} \cmidrule(lr){7-7} \cmidrule(lr){8-9} \cmidrule(lr){10-11}
& \textbf{CLIP}
& \textbf{CLIPSeg}
& \textbf{DINO}
& \textbf{YOLO}
& \textbf{MRCNN}
& \textbf{FRCNN}
& \textbf{InternVL3}
& \textbf{Qwen2.5}
& \textbf{InternVL3}
& \textbf{Qwen2.5} \\
\midrule
MLIC++ - MS-SSIM   
& -12.6 & -26.2 & 0.5 & -5.9 & -4.8 & -22.3 & 12.2 & 12.2 & -30.4 & 5.9 \\

UG-ICM~\cite{yin2025unified}           
& -29.4 & -7.4 & -27.0 & -23.3 & -12.3 & -42.6 & -11.0 & -16.6 & -27.5 & -24.1 \\

MLIC++ - LPIPS     
& -46.8 & -39.1 & -37.7 & 54.0 & -37.8 & -66.3 & -20.8 & -20.8 & -27.4 & \textcolor{blue}{-37.0} \\

MLIC++ - CLIPScore 
& \textcolor{blue}{-68.4} & \textcolor{blue}{-69.8} & \textcolor{blue}{-55.6} & \textcolor{blue}{-46.2} & \textcolor{blue}{-47.8} & \textcolor{blue}{-70.2}
& \textcolor{blue}{-41.8} & \textcolor{blue}{-30.4} & \textcolor{blue}{-43.7} & -36.1 \\

\textbf{MLIC++ - ML-CLIPSim}
& \textbf{\textcolor{red}{-76.1}} & \textbf{\textcolor{red}{-79.9}} & \textbf{\textcolor{red}{-62.9}} & \textbf{\textcolor{red}{-57.3}} & \textbf{\textcolor{red}{-57.5}} & \textbf{\textcolor{red}{-80.7}}
& \textbf{\textcolor{red}{-57.9}} & \textbf{\textcolor{red}{-31.3}} & \textbf{\textcolor{red}{-61.8}} & \textbf{\textcolor{red}{-57.5}} \\
\bottomrule
\end{tabular}
}
% \vspace{-6pt}
\end{table}
We train MLIC++ codecs using ML-CLIPSim as the distortion term and compare against MSE/PSNR, MS-SSIM, LPIPS, CLIPScore, and UG-ICM~\cite{yin2025unified}, a strong task-agnostic machine-oriented compression baseline.
Compared with earlier ICM methods that are often designed for specific tasks or require task-dependent adaptation, UG-ICM provides a more general and scalable solution, serving as a strong and representative baseline in our comparison.

\noindent\textbf{Implementation Details.}
For learned image compression experiments, we fine-tune pretrained MSE-optimized MLIC++ codecs on ImageNet using ML-CLIPSim as the distortion term, while keeping the ML-CLIPSim network fixed. We train with Adam for 5 epochs using $256\times256$ crops, batch size 8, learning rate $10^{-4}$, and $\lambda\in\{0.6,2,6,10,18\}$ to obtain different rate points. All objectives are evaluated under the same codec architecture and downstream evaluation protocol.

The main rate--task results are shown in Fig.~\ref{fig:codec_all_curves}, covering ImageNet classification, PASCAL VOC object detection, and COCO downstream tasks. These experiments focus on machine-oriented performance, which is the primary objective of our method.

Fig.~\ref{fig:codec_all_curves} shows rate--task curves on ImageNet, VOC, and COCO using MLIC++ codecs trained with different distortion objectives. ML-CLIPSim consistently achieves the best rate--task trade-off across classification, detection, and segmentation tasks. Table~\ref{tab:bd_rate_dataset} further confirms this trend, with ML-CLIPSim obtaining the best BD-Rate on all evaluated tasks, including $-76.1\%$ on ImageNet classification, $-79.9\%$ on COCO CLIPSeg, and $-80.7\%$ on VOC detection. 
ML-CLIPSim also consistently outperforms the strong UG-ICM baseline across all evaluated settings. While UG-ICM already achieves strong cross-task generalization through CLIP-based semantic supervision~\cite{yin2025unified}, our method further improves performance by learning a structured proxy of machine utility from multi-model predictive consistency, leading to more consistent gains across tasks.

We note that earlier ICM methods are not included in this comparison, as they are typically designed for single-task or task-adaptive scenarios and are not directly comparable in our unified multi-task evaluation setting.
Compared with CLIPScore, ML-CLIPSim achieves more consistent gains across downstream tasks, suggesting the benefit of multi-layer token representations for machine-oriented quality modeling.

We further evaluate VLMs (InternVL3-1B and Qwen2.5-VL-3B-Instruct)~\cite{chen2024internvl,wang2024qwen2} on SEEDBench\_IMG and POPE~\cite{li2023seed,li2023evaluating}. 
As shown in Fig.~\ref{fig:vlm_main} and Table~\ref{tab:bd_rate_dataset}, ML-CLIPSim-trained MLIC++ codecs achieve the best BD-Rate on both POPE and SeedBench with both VLM backbones.
We observe slight non-monotonicity in rate--task curves, which is common in machine-oriented evaluation.
% We also evaluate reconstruction quality on Kodak (Fig.~\ref{fig:kodak_results}).
We additionally evaluate reconstruction quality on Kodak, with detailed results and additional experiments on Hyperprior-based codecs provided in the supplementary material.

% \begin{figure}[t]
% \centering
% \setlength{\tabcolsep}{0pt}
% \begin{tabular}{@{}c@{\hspace{0.02\columnwidth}}c@{}}
% \includegraphics[width=.45\linewidth]{paper_figures_combined_scheme1/kodak_psnr_combined_scheme1.pdf} &
% \includegraphics[width=.45\linewidth]{paper_figures_combined_scheme1/kodak_msssim_combined_scheme1.pdf} \\
% \end{tabular}
% \caption{Rate--distortion comparisons on Kodak under PSNR and MS-SSIM. ML-CLIPSim maintains competitive reconstruction quality while optimizing machine-oriented utility.}
% \label{fig:kodak_results}
% \end{figure}

% ------------------------------------------------------------
\subsection{Ablation Study}
% \begin{table*}[t]
% \centering
% \caption{Correlation on human-IQA benchmarks and PCMP. Higher is better.}
% \label{tab:main_allcorr}
% \small
% \setlength{\tabcolsep}{3pt}
% \renewcommand{\arraystretch}{1.1}
% \resizebox{.8\textwidth}{!}{%
% \begin{tabular}{l | ccc | ccc | cccc}
% \toprule
% \multirow{2}{*}{Metric}
% & \multicolumn{3}{c}{TID2013}
% & \multicolumn{3}{c}{LIVE}
% & \multicolumn{4}{c}{Proposed (PCMP)} \\
% \cmidrule(lr){2-4}\cmidrule(lr){5-7}\cmidrule(lr){8-11}
% & SRCC$\uparrow$ & KRCC$\uparrow$ & PLCC$\uparrow$
% & SRCC$\uparrow$ & KRCC$\uparrow$ & PLCC$\uparrow$
% & SRCC$\uparrow$ & KRCC$\uparrow$ & PLCC$\uparrow$ & ACC$\uparrow$  \\
% \midrule
%
% CLIPScore (global only)
% & 0.744 & 0.552 & 0.794
% & 0.892 & 0.712 & 0.903
% & 0.695 & 0.515 & 0.610 & 0.803 \\
%
% ML-CLIPSim token only
% &0.709 &0.532 &0.782
% &0.925 &0.754 &0.929
% &0.793 &0.619 &0.735 &0.827 \\
%
% ML-CLIPSim full w/o group
% &0.748 &0.561 &0.805
% &0.931  &0.765  &0.935
% &0.789 &0.613 &0.738 &0.826\\
%
% ML-CLIPSim equal layer weights
% &0.748 &0.562 &0.805
% &0.932 &0.765  &0.935
% &0.787 &0.611 &0.736 &0.826 \\
%
% \midrule
% ML-CLIPSim (Ours, hard)
% &0.752 &0.565 &0.807
% &0.932 &0.766 &0.936
% &0.788 &0.612 &0.736 &0.826\\
% ML-CLIPSim (Ours, soft)
% & 0.734 & 0.551 & 0.798
% & 0.929 & 0.760 & 0.934
% & \textbf{0.794} & \textbf{0.618} & \textbf{0.736} & \textbf{0.828} \\
%
% \bottomrule
% \end{tabular}%
% }
% \end{table*}

\begin{table}[t]
\centering
\caption{Ablation results on PCMP. Higher is better.}
\label{tab:main_allcorr}
\small
\setlength{\tabcolsep}{3pt}
\renewcommand{\arraystretch}{1.1}
\resizebox{.85\columnwidth}{!}{%
\begin{tabular}{l | cccc}
\toprule
Metric & SRCC$\uparrow$ & KRCC$\uparrow$ & PLCC$\uparrow$ & ACC$\uparrow$ \\
\midrule
global (CLIPScore)
& 0.695 & 0.515 & 0.610 & 0.803 \\

global + multi-layer token + uniform weights
& 0.787 & 0.611 & 0.736 & 0.826 \\

global + multi-layer token + learned layer weights
& 0.789 & 0.613 & 0.738 & 0.826 \\

multi-layer token + grouped layers
& 0.793 & 0.618 & 0.735 & 0.827 \\

% \midrule
global + multi-layer token + grouped layers (hard label)
& 0.788 & 0.612 & 0.736 & 0.826 \\
% \midrule
\textbf{ML-CLIPSim (ours)}
& {0.794} & {0.618} & {0.736} & {0.828} \\
\bottomrule
\end{tabular}%
}
% \vspace{-6pt}
\end{table}

Table~\ref{tab:main_allcorr} summarizes the ablation results on PCMP. 
Compared with global CLIP similarity alone, incorporating multi-layer token representations substantially improves correlation with machine-oriented preference labels. 
We further observe that grouped layer aggregation and soft pairwise supervision provide consistent gains, indicating that both structured cross-layer modeling and soft predictive-consistency supervision contribute to improved estimation of latent machine utility.

% We ablate key design choices in ML-CLIPSim and the supervision protocol, including token-only vs.\ token+global, layer count $K$, PSNR-matching on/off, and tie filtering.
% We additionally evaluate on human IQA benchmarks (e.g., TID2013 and KADID-10k) to analyze the human--machine trade-off and motivate the default last-$K$ design.
% Due to space constraints, detailed ablation tables and human-IQA correlations are included in the supplementary.
\section{Conclusion}
We presented the Predictive Consistency Dataset for Machine Perception (PCMP), a predictive-consistency dataset with continuous pairwise labels derived from multi-model voting.
We also introduced ML-CLIPSim, a learnable multi-layer CLIP similarity metric for machine-oriented image quality assessment.
By training lightweight modules over frozen CLIP representations with downstream consistency supervision, we obtain a differentiable and scalable proxy for machine-oriented image quality.
When used as a distortion term for learned image compression, the proposed metric improves downstream task performance at comparable bitrate, and our ablation study shows that multi-layer token modeling, structured layer aggregation, and soft pairwise supervision are important for estimating latent machine utility.
We hope the dataset and metric will facilitate broader research on machine-oriented image quality assessment, utility-driven compression, and machine-centric visual communication.

\bibliography{egbib}
\end{document}